# 60 GHz Wi-Fi As A Tractor-Trailer Wireless Harness

Ahmed Elhadeedy
*Department of Systems Engineering*
Colorado State University
PA, USA
aelhadee@gmail.com

Jeremy Daily
*Department of Systems Engineering*
Colorado State University
CO, USA
Jeremy.Daily@colostate.edu

*Abstract*—Reverse driving a truck is a challenging task for human drivers and self-driving software due to the lack for sensors on the trailer. Self-driving and conventional trucks have an increasing need to replace the legacy communication channels between the truck and the trailer to accommodate bandwidth and latency requirements when more sensors and features are added to the trailer to support driver assist features or self-driving functions, in addition to the need of automating the tractor-trailer hitching and unhitching, which is a complex process when using wires and connectors for communication between the truck and the trailer. In this paper, we address using a wireless harness between the tractor and the trailer based on Wi-Fi, in addition to discussing using Named Data networking protocol for communication between the truck and the trailer including handling interest and data packets. A Testbed is used to evaluate communicating different data types from one device to three devices over 802.11ac and it indicated a stable communication performance when Named Data Networking and Data Distribution Service were used. Using a wireless harness will ease the automation of trailer hitching and unhitching process and will eliminate the need for communication wires or connectors between the tractor and the trailers, therefore, reducing the complexity of the process.

## I. Introduction

Driverless and conventional trucks are becoming more capable with the addition of autonomy software and Advanced Driver Assistance Systems (ADAS) systems that bring different sensors to the truck. One of the challenging maneuvers for human and software drivers in semi-trucks is reverse driving due to the lack of rear facing sensors on the trailer to aid the driver while backing up. Legacy communication channels between the truck and the trailer has basic bitrate and cannot support sensors data transfer or meet the latency and bandwidth requirements for the new autonomy systems. There have been proposals to use Ethernet between the trailer and the truck to support the new requirements, in addition to the automation of trailer hitching and unhitching process. Automation of the tractor-trailer physical hitching and unhitching will include plugging and unplugging data and control wires and connectors from the tractor to the trailer, which is a complex process to automate that requires the connectors to mate properly to fully plug in, which, for example, could be done using a robotic arm on the tractor or positioning the connectors in a precise location in each of the tractor and the trailer. It's expected that driverless trucks will not need a human during its mission including pickup and drop off.

Wireless harness is one of the concepts that can meet the communication requirements, bandwidth and at the same time greatly help the automation of hitching and unhitching process since no wires or connectors will be needed and the pairing process between the truck Electronic Control Units (ECU) and the trailer ECU will be securely automated. Wireless harness has been proposed previously for vehicle in-cabin communication between different ECUs but has not been addressed for truck and trailer communication. Wireless harness leverages variety of technologies and has been tested for in-cabin communication with promising results, technologies such as Impulse Radio Ultra-Wideband (IR-UWB) [1], Millimeter-Wave [2], ZigBee [3], IEEE 802.11ad [4] and IEEE 802.15.1 [5]. IEEE802.11ad appears to be the most appropriate technology to use since it can support the requirements of a gigabit or multi-gig networking between the truck and the trailer and at the same time, it's a well-established technology with more development resources. IEEE802.11ad could be used between the trailer ECU and the truck ECU as shown in Fig. 1Fig. 1 where an Ethernet-Wi-Fi bridge could be used on both sides to enable communication between the trailer ECU and other ECUs on the truck side whether they are connected to an Ethernet Network using a switch, a Controller Area Network (CAN) bus which is managed by the truck gateway to enable two-way communication between the remote ECU and local ECUs from heterogenous networks.

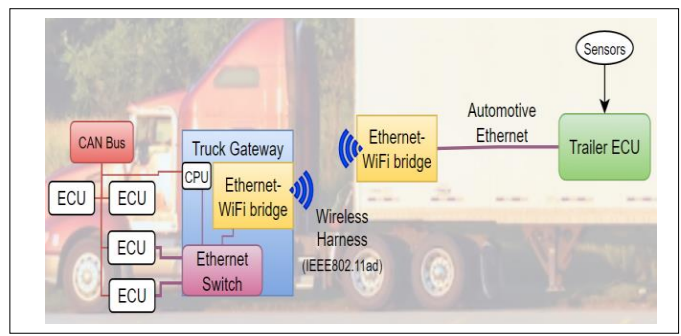

Fig. 1. Concept of using IEEE802.11ad as a wireless harness for truck and trailer communication.

Using the wireless harness for communication between the truck and the trailer will eliminate the need for additional data wires or connectors since the heterogenous data could be digitized and wrapped in Wi-Fi frames including CAN, CAN Flexible Data (FD), sensors data or diagnostics data. This newly introduced intra-communication will require a new networking approach to manage the different types of data exchanged including integration with existing native truck network and security such as data confidentiality, integrity, and authenticity since the attack surface will increase. One of new networking protocols that has security by design and has been tested in the context of automotive communication with positive results is Named Data Networking (NDN) [6] [7] [8]. We previously discussed using Ethernet, NDN and secure automated pairing in [9].

In this paper, we discuss NDN when used for truck and trailer communication and tested two different networking approaches over a wireless network to evaluate the performance and the stability of the communication when automotive-like data is used. The first approach is Data Distribution Service (DDS) which is a well-known networking protocol that used publisher and subscriber model, and the second approach is NDN which is similar to DDS.

## II. NAMED DATA NETWORKING

Named data networking is a data-centric networking protocol that uses data packets and interest packets for communication where each data type has a name that is being used for communication instead of an Internet Protocol address (IP). Data packet contains the name of the data, the content, meta data such as freshness period, and the signature. Similarly, for the interest packet contains the name of the data, meta data such as interest lifetime and an optional signature. NDN forwarding Daemon (NFD) [10] handles the routing and forwarding the interests packets and data packets using the forwarding plane [11] based on the predefined networks interfaces. Let's say ECU1 is one the truck with NFD1 and ECU2 is on the trailer with NFD2. The application in ECU1 needs sensor data from the trailer that is named */trailer/sensor1* so it generates an interest packet using that name. NFD1 checks the content Store (CS) of ECU1 where the received data is cached, if not match, it checks for the interest packet in Pending Interest Table (PIT), it there is a match it adds the new interface to the table, if not, NFD1 checks in Forwarding Information Base (FIB) to identify the forwarding route for the interest packet based on the prefix of the name and then it will determine the next hop. The interest packet is delivered now to NFD2 in ECU2 where it checks CS2 for the data that matches the new interest packet, if not match, it will check PIT2 and if not match it will add a pending interest packet to the table. NFD2 will check FIB2 for the route to determine the next hop, and the interest packet will be forwarded to ECU2 application to get the data packet as show in Fig. 2, otherwise, it will forward it to the next node as defined in FIB2. The app can validate the interest packet using the validation key and check other parameters such as freshness period of the interest.

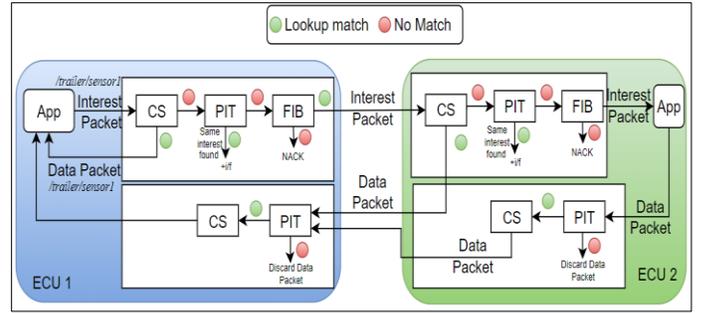

Fig. 2. Interest packet and data packet example communication when ECU1 is requesting data from ECU2 over NDN.

The Application within ECU2 will respond with the data packet so NFD2 will check if the interest is still valid within PIT2, if so, it can cache the data within CS2 before forwarding the data packet to ECU1. Once the data packet is forwarded to NFD1, it will check PIT1, if there is a match, it will cache the data in CS1 and route the data to the application or as defined in FIB1.
The data packet will be signed using ECU1, the producer, private key and ECU1, the consumer, can validate the data packet using the validation or the public key. Similarly, interest packets could be signed and validated the same way assuming ECU1 and ECU2 did the key exchange for authentication.

## III. SYSTEM REQUIREMENTS

The overall requirements of the communication system between the truck and the trailer are as shown in TABLE I. :

TABLE I. SYSTEM REQUIREMENTS

| ID  | Requirement |
| --- | --- |
| R1  | The system shall support automotive communication requirements |
| R2  | The system shall be integrated with the existing platforms hardware |
| R3  | The system shall be able to communicate with heterogenous automotive networks simultaneously |
| R4  | Th system shall support data integrity, authenticity, and confidentiality |
| R5  | The system shall support communication timing requirements such as CAN signals timing of 20 milliseconds |
| R6  | The system shall support at least 1 gigabit bandwidth |
| R7  | The wireless harness shall support automated pairing between trailer ECU and the tractor ECUs or other authorized devices |
| R8  | The wireless harness shall support diagnostics and software updates access |
| R9  | The wireless harness shall have high reliability and stability like the physical connections |
| R10 | The wireless harness shall be stable at normal driving conditions and maneuvers |

## IV. DATA MANAGEMENT

Within NDN, different data is given different name such as */trailer/sensor1*, */trailer/sensor2*, */trailer/J1939, and /trailer/CANFD*. NDN Data packets are 8800 bytes in size which allows for a bigger payload. The payload of the data packets could be managed further to transmit different data simultaneously such as multiple CAN frames or sensors data with CAN frames attached to it as an annotation.

The trailer ECU will collect the data from different sources, format them based on their standard specification and the intended destination at the tractor side, construct them using a software multiplexer in one construct $P$ as follows:

$$P = (c, s, A, a) \quad (1)$$

$P$ is the hybrid construct that contains $m$ frames of heterogenous vehicle network protocols or sensor bytes, $c$ is the total size of $P$, $s$ = timestamp for the construct $P$, $A$ is the hybrid payload that contains automotive protocol frames, sensor bytes, or any type of data that will be transmitted to truck communication buses such as J1939, CAN FD, radar CAN and Ethernet, or LiDAR data bytes and $a$ is an authentication tag, if not included by default (i.e., other protocols), that results from the encryption of $c$, $s$ and $A$.

$$A = [r_i \quad t_i \quad p_i \quad l_i \quad f_i] \quad (2)$$

$r$ is the transmission priority number assigned to each protocol or data frame, $t$ is timestamp for each data type, $p$ is the protocol definition (e.g., $p_0$ = Sensor data bytes, $p_1$ = J1939 frame, $p_1$ = CAN FD frame, $p_2$ = CAN FD frame and $p_m$ is any other automotive data or protocol that could be packed as a part of $A$), $l$ = total length of each data type (e.g., number of bytes of the sensor data or the CAN frame), $f$ is the actual CAN or sensors bytes that will be transmitted to the AT communication bus or to the AT ECUs and $i$ is the index for each unique data frame, $i = [0,1,\cdots,m]$, where $m$ is the maximum number of frames within $A$.

$$A = \begin{bmatrix} r_0 & t_0 & p_0 & l_0 & f_0 \\ r_1 & t_1 & p_1 & l_1 & f_1 \\ \vdots & \vdots & \vdots & \vdots & \vdots \\ r_m & t_m & p_m & l_m & f_m \end{bmatrix} \quad (3)$$

$A$ could contain multiple data types such as sensor data ($i = 0$), LiDAR data packet ($i = 1$) along with CAN FD frame 1 ($i = 2$), CAN FD frame 2 ($i = 3$), J1939 CAN frame ($i = 4 = m$) frames or any other data type. The size of $A$ will vary depending on the used networking protocol and the maximum allowed payload, in the case of NDN, the maximum size for the data packet is 8800 bytes.

A different $P$ construct could be used for each data type. For example, $P_v$ for one video frame ($i = m = 0$ in $A$), $P_l$ for a LiDAR data packet ($i = m = 0$ in $A$) and $P_s$ for AT serial bus data which will contain multiple frames, first CAN FD frame is for AT CAN FD channel 1 ($i = 0$), second CAN FD frame is for AT CAN FD channel 2 ($i = 1$) and third CAN frame is for the AT J1939 bus ($i = m = 3$).

## V. TEST AND EVALUATION

In this section, NDN is compared with DDS when used over a wireless medium and automotive-like data is being transmitted.

### A. Testbed

The test setup includes a PC is the data producer (e.g., trailer ECU) communicating over a Wi-Fi connection with three receivers via a router. The receivers are wired to the router with Ethernet cable. The receivers (e.g., truck ECUs) are Raspberry Pis (RPi) running Ubuntu Server 21.10 OS. The Wi-Fi network is 802.11ac, 5.745GHz, transmission power of 22 dBm and signal level of -30 dBm. Using the 60GHz Wi-Fi on a trailer is part of the future work. 802.11ac was used for networking evaluation and Wi-Fi link stability when used in this setup and data types.

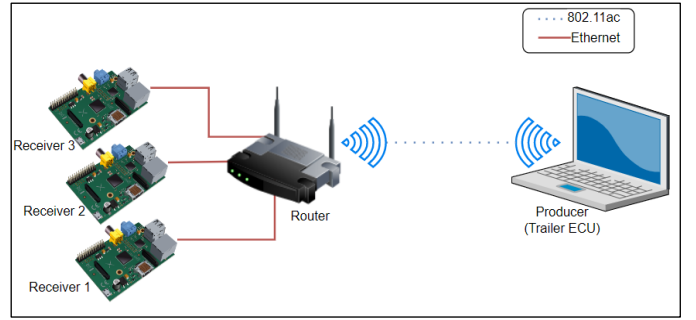

Fig. 3. Testbed used for evaluating the networking protocols.

### B. Test Configuration and Method

The PC is hosting three producers' scripts, one for each data type. Assuming we have three different types of data as follows: lidar data, cam data and CAN data. The lidar data is 1600 bytes transmitted every 5 milliseconds (*ms*), the CAN data is 160 bytes transmitted every 8 *ms* and the cam data is 4000 bytes transmitted every 20 *ms*. Each RPi is receiving only one data type hosting one receiving script. In DDS, three topics were created and within each there is string data type, and each receiver can subscriber to one data type.

*1) NDN configuration*

Each data type is given a name as follows: */trailer/cam*, */trailer/lidar*, */trailer/can*. The test requires defining interfaces (faces) between the producer and the consumers as shown in TABLE II.

TABLE II.  FACES DEFINITION

| Node | Local Address | Route |
| --- | --- | --- |
| PC | Face1 to RPi1: udp://192.168.10.11 | */trailer/lidar* via face1 |
| | Face2 to RPi2: udp://192.168.10.12 | */trailer/can* via face2 |
| | Face3 to RPi3: udp://192.168.10.13 | */trailer/cam* via face3 |
| RPi1 | Face1 to PC: udp://192.168.10.33 | */trailer/lidar* via face1 |
| RPi2 | Face1 to PC: udp://192.168.10.33 | */trailer/can* via face1 |
| RPi3 | Face1 to PC: Udp://192.168.10.33 | */trailer/cam* via face1 |

## 2) Test Method

DDS Real-Time Publisher-Subscriber (RTPS), NDN over Transmission Control Protocol (TCP) and NDN over User Datagram Protocol (UDP) were tested separately using the testbed. DDS was limited to a string variable only as the payload where NDN was tested with serializing bytes, referred to as (B) as well as a string variable, referred to as (S). In the case of the string variable, a string variable is generated at the beginning of the test and then randomized for each data packet transmission and the random bytes were generated with each data packet transmission. In the case of NDN, Each receiver will be simultaneously sending an interest packet with a different data name to the producer to get a new data packet. Similarly, DDS subscribers getting the data published simultaneously.

## C. Test Results and Discussion

The two performance parameters captured were latency and core CPU consumption. Latency was calculated as the time difference between each received packet at each receiver. CPU utilization percentage was captured at the data producer and each of the consumers as well. The dashed red line in the latency results is periodic transmission time or the hardcoded delay. Wi-Fi Encryption and data packets signing are not in the scope of this paper and will be addressed as part of the future work.

### 1) Latency

Latency of received CAN packets looked overall similar between all the approaches as shown in Fig. 4 with DDS RTPS slightly better performance.

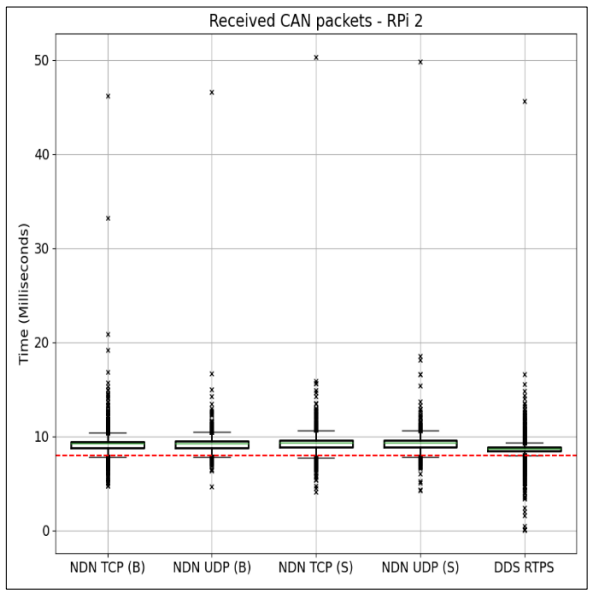

Fig. 4. Latency comparison for */trailer/can* data at the receiver

For the latency when lidar packets where transmitted, it's an increased load of 1600 bytes so the difference in performance started to appear especially when serializing bytes compared to serializing strings as shown in Fig. 5. NDN (B) is performing better due to the less encoding that needs to be done on the payload. NDN (S) and DDS used strings and they had similar performance due to the additional encoding done at the producer in both cases. However, DDS used JSON encoding where NDN used Type-Length-Value (TLV). JSON is more efficient for strings when compared with TLV with strings since TLV encodes the data in binary format, therefore, more bytes will be required when encoding strings in the case of NDN and it will cause the latency performance of the application to degrade when serializing strings. NDN

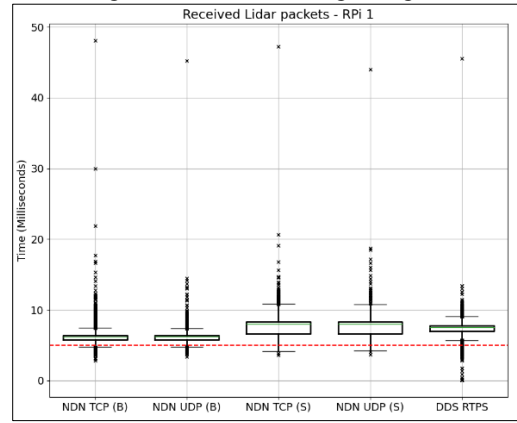

Fig. 5. Latency comparison for */trailer/lidar* data at the receiver

Similarly, for */trailer/cam,* NDN and DDS serializing strings show similar performance showing similar mean latency with DDS performing slightly better. NDN serializing bytes in shown to be the most efficient with bigger payload when serializing bytes.

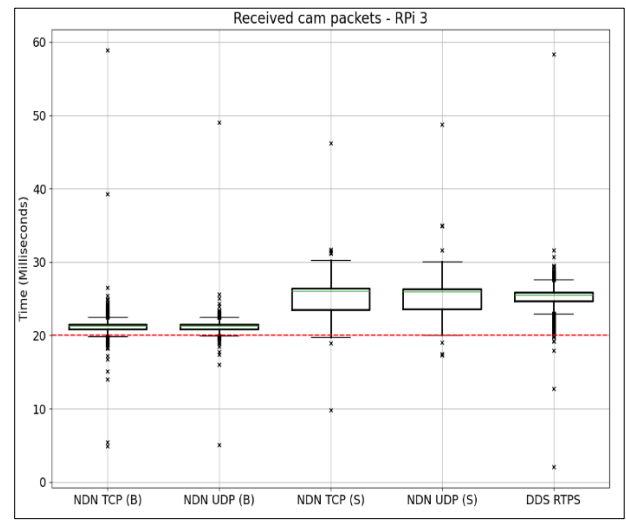

Fig. 6. Latency comparison for */trailer/cam* data at the receiver

### 2) Core CPU Utilization

Core CPU utilization percentage was recorded at the producer for each producing script and on each receiver. NFD was also added for NDN at each of the producer and the receiver.

#### a) CPU Utilization At The Transmitter

There was no high difference between the CPU utilization percentage when comparing the scripts of each data type as shown in Fig. 7, for example, lidar over NDN TCP is 2-3% higher than Lidar over DDS. Overall, they have similar utilization and tolerable differences. It was also shown that NFD had a low CPU utilization in both cases.

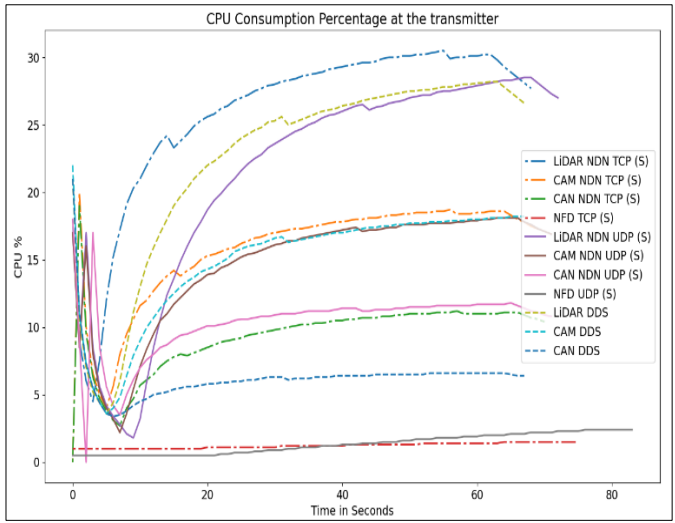

Fig. 7. CPU Utilization percentage at the producer for each script

*b) CPU Utilization At The Receiver*

Fig. 8 shows the CPU utilization percentage for each script at each receiver. Overall, the three networking approaches had no high difference in CPU utilization, however, NDN TCP had a slightly higher utilization due to the continuous transmission of interest packets to the produces. In the case of the NDN over TCP, its response with an ACK in addition to the interest packet.

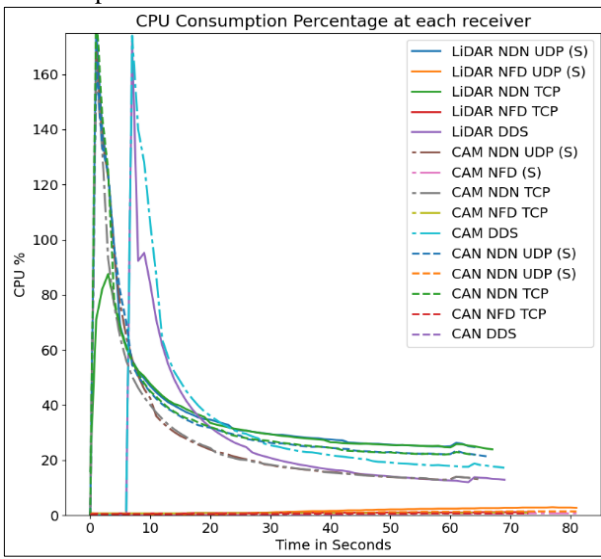

Fig. 8. CPU Utilization percentage for each script at each receiver (RPi)

Within this setup, NDN had a similar performance or slightly less when compared to a well-established protocol such as DDS when serializing strings over Wi-Fi. Additionally, the test did not show abnormal number of latency spikes or Wi-Fi lagging when sending data packets at that fast rate and the number of latency outliers is limited. NDN also is shown to be efficient when used to serialize bytes over Wi-Fi and Ethernet especially for bigger packets such as the case above of 1600 bytes and 4000 bytes.

## VI. CONCLUSION AND FUTURE WORK

This paper proposes using the 60 GHz Wi-Fi wireless harness for communication between the truck and the trailer to combat bandwidth and timing limitations in the existing truck-trailer communication and at the same time support the automation process of coupling and uncoupling a truck with a trailer by enabling the automated pairing between truck ECU and trailer ECU and eliminate the need for data wires. Using NDN for truck-trailer communication was also discussed and how interest and data packets will be handled between the two ECUs. A testbed is used to evaluate NDN and DDS over 802.11ac link and the test indicated that NDN and DDS in the case of serializing strings had similar performance and CPU utilization with DDS being slightly better due to the differences in encoding used. Additionally, serializing bytes over NDN is shown to be the most efficient approach when compared NDN and DDS serializing strings.

Future work will include testing the system using an IEEE802.11ad transmitter and a receiver on a trailer with different configurations to evaluate the wireless link and if possible, on an actual semi-truck to evaluate the performance under noisy conditions and the influence of different driving conditions and maneuvers on the reliability of the wireless harness between the truck and the trailer.